# Vision Transformers and Bi-LSTM for Alzheimer's Disease Diagnosis from 3D MRI

Taymaz Akan
*Department of Medicine*
Louisiana State University Health Sciences Center at Shreveport
Shreveport, USA
taymaz.akan@lsuhs.edu

Sait Alp
*Department of Computer Engineering*
Erzurum Technical University
Erzurum, Turkey
saeid.agahian@erzurum.edu.tr

Mohammad A. N Bhuiyanb
*Department of Medicine*
Louisiana State University Health Sciences Center at Shreveport
Shreveport, USA
nobel.bhuiyan@lsuhs.edu

*Abstract*— Alzheimer's is a brain disease that gets worse over time and affects memory, thinking, and behavior. Alzheimer's disease (AD) can be treated and managed if it is diagnosed early, which can slow the progression of symptoms and improve quality of life. In this study, we suggested using the Visual Transformer (ViT) and bi-LSTM to process MRI images for diagnosing Alzheimer's disease. We used ViT to extract features from the MRI and then map them to a feature sequence. Then, we used Bi-LSTM sequence modeling to keep the interdependencies between related features. In addition, we evaluated the performance of the proposed model for the binary classification of AD patients using data from the Alzheimer's Disease Neuroimaging Initiative (ADNI). Finally, we evaluated our method against other deep learning models in the literature. The proposed method performs well in terms of accuracy, precision, F-score, and recall for the diagnosis of AD.

## I. INTRODUCTION

AD is a neurodegenerative disorder caused by abnormal protein deposits in the brain, causing nerve cells to degenerate and eventually die. This leads to diminished cognitive function, altered mood, and behavior [1], [2]. Alzheimer's disease has no known cure, but treatments can manage symptoms and improve life. Common symptoms include memory loss, difficulty with tasks, language difficulties, disorientation, poor judgment, abstract thought issues, object misplacement, mood changes, and motivation loss. Alzheimer's disease begins in the preclinical stage, where changes in the brain and blood are not visible. It is believed that AD begins at least 20 years before symptoms appear. The second phase is mild cognitive impairment (MCI), and dementia is the final stage, where memory, thinking, cognitive abilities, and behavioral symptoms impair daily life [3]. Early diagnosis allows for early treatment and management, slowing the progression of symptoms and improving the quality of life for Alzheimer's patients. Non-invasive medical tests, such as X-rays, CT scans, and MRI scans, can also help in early detection and management.

Alzheimer's disease detection can also be classified as either invasive or non-invasive. Obtaining data from the interior of the patient's body through invasive procedures, such as a lumbar puncture or blood extraction, is necessary for invasive approaches. These invasive techniques aim to identify possible biomarkers that are accurate indicators of Alzheimer's disease. Most of them aren't completely risk-free for the patient and often cause a great deal of discomfort, if not outright pain. In contrast, non-invasive diagnostics are risk-free and more practical because the diagnostic procedure does not involve breaking the skin or entering the body in any way. These tests are usually painless and carry a minimal risk of complications. Consequently, they are more convenient during the diagnostic process [10], [11]. Non-invasive medical tests can include imaging tests such as X-rays, CT scans, and MRI scans. Brain MRI has benefits over CT, including the absence of bone hardening artifacts (especially bothersome in the medial temporal lobe regions), the possibility of obtaining oblique and transverse cuts, and the ability to differentiate white and gray matter in the medial temporal lobe structures [12].

Computer-aided diagnosis (CAD) systems have made AD classification easier for neurologists. Conventional and deep learning-based CAD systems exists. Deep learning algorithms require minimal image pre-processing and can automatically infer optimal data representations without feature selection, unlike conventional four-stage pipelines [4]. Deep-based architectures like CNN [5], RNN [6], and Transformers [7] are widely used for medical image analysis, especially in 2D and 3D ultrasound and MRI images. 3D MR brain images are stacked 2D data slices, making 3D-CNN deep models impossible.

The Transformer architecture, which dominates natural language processing [8], has gained popularity in computer vision due to its impressive results in tasks like image classification, object detection, and semantic segmentation. ViT, based on Transformers, has been applied to images with minimal modifications and has shown superior performance in many computer-vision tasks, making it a viable alternative to CNNs as a network architecture [9]. CNNs gradually collect features from local to global using convolutional layers. ViT, on the other hand, uses a multi-headed self-attention mechanism to capture long-range dependencies, allowing the model to focus on all elements in the input sequence. This makes ViT ideal for brain imaging analysis due to its ability to accurately capture interdependencies between dispersed brain regions [10]. However, the training set for MRI is too small, necessitating massive data to train a ViT-based model from scratch. To avoid missing inductive bias, transfer-learning-based models are necessary. ViT has so far extracted the features of each slice separately. However, maintaining relationships between the slices is inevitable. Therefore, a Recurrent Neural Network (RNN) like LSTM is required to process sequential data.

This study used a pre-trained ViT architecture to extract features from each 2D slice. The brain's feature sequences will be encoded by a bi-LSTM model for binary classification. We tested the proposed model on ADNI datasets against other deep learning models in the literature.

## II. RELATED WORKS

The superior performance of convolutional neural networks (CNN) in computer vision has actually been proven

[11]. They have proven to be very practical in many computer vision tasks, including those involving image classification [39], object detection and tracking [12], [13], and semantic segmentation [14]. In addition, medical image analysis has made extensive use of these CNN-based architectures. They have been applied to 2D and 3D ultrasound and MRI images [15]. CNNs are the deep models that are utilized the most frequently to detect AD [15]–[17].

Since an MR brain scan is 3D data consisting of stacked 2D slices, a model with the capability of describing the whole brain at once is required. Hence, the most straightforward method for this purpose is a 3DCNN model to learn spatiotemporal features, which is not possible with a 2DCNN.

[18] suggested residual and plain 3D-CNN architectures for AD multi-classification on 3D structural MRI brain scans from the Alzheimer's Disease National Initiative (ADNI). Moreover, [19] suggested using a 3D-CNN that can learn generic features that capture AD biomarkers and adapt to different domain datasets to predict AD. The 3D-CNN is based on a 3D convolutional autoencoder that has already been trained to recognize changes in the shape of the brain from structural MRI scans. Then, the fully connected upper layers of the 3D-CNN are fine-tuned for each task-specific AD classification. Also, in [20], an end-to-end 3D-CNN model was proposed. This model would use MRI voxels from the whole brain at once to capture both the subtle local brain details and the more obvious global MRI features. Finally, the ADNI dataset is classified into three classes (AD, MCI, and NC) using a learning transfer strategy for MCI classification.

Because it requires a large number of parameters and a high amount of computation, the 3D model does not have the capability to construct deep models [15]. This research chooses either 2D or 3D CNN models, so there are trade-offs between the two approaches to learning representations for 3D medical images [21]. Annotation-efficient deep learning with limited data often relies on transfer learning, which uses pretrained weights from large-scale datasets (e.g., ImageNet [22]) [21]. CNNs were initially applied to 2D images for decision-making purposes. For this reason, slice-based techniques are used to analyze MRI scans that split 3D neuroimages into 2D slices. When using slide-based methods to classify 3D-MRI scans, it's crucial to consider the spatial features in each slice and how those features depend on each other across slices (i.e., voxels' dependencies). Pre-trained off-the-shelf 2D-CNN is capable of extracting special features from each slice. Deep models used for sequence modeling, like 1D-CNN, RNN-based models, and their derivatives, can subsequently perform slice dependency modeling. Furthermore, temporal CNNs (TCNs) can simultaneously extract features and classify sequences [23].

The Transformer architecture, which dominates natural language processing, has limited applications in computer vision [8]. Attention is often used alongside convolutional networks or to replace parts of convolutional networks while maintaining their overall structure. However, convolutional architectures remain the standard in computer vision. Several works aim to integrate CNN-like architectures with self-attention, sometimes replacing them entirely. Vision Transformer (ViT) has gained popularity due to its impressive results in various computer vision tasks, including image classification, object detection, and semantic segmentation.

ViT takes inspiration from Transformers' success in NLP and applies a standard Transformer to images with minimal modifications. Unlike CNNs, ViT can extract features across the entire image without compromising image resolution, preventing spatial loss from information skipping. The self-attention module is crucial in ViT, gathering information from all locations in an image to extract features. The entire image region is processed without changing resolution, and local image information is stored in patch features [24].

[10] proposed a model using ViT for binary (AD/CN) classification on MRI. They split the standardized 3D MRI coronal slices into 2D slices. Later, only the middle 75 slices were considered the model input. Every slice was treated as a new sample and given the same label as the subject it came from. They used a modified patch-wise embedding (Convolution operation with overlapping) layer to adapt the pre-trained ViT model to classify each slice independently. Training the modified ViT was performed with a pair of new samples and corresponding labels. The slice vote results determine each subject's final prediction. The final prediction for each MRI was generated by voting among the predicted labels of slices. As mentioned before, maintaining inter-slice relationships is inevitable. However, this method suffers from keeping temporal information between the slices..

[25] came up with the idea of using multiple vision transformer networks in AD classification. They take 15 2D slices from the center of the coronal plane and put them next to each other. Every slice was treated as a new sample and given the same label as the subject it came from. The ViT model was applied to classify each slice independently. Subsequently, another transformer, namely GFNet [26], was applied to classify each slice again. The GFNet uses the same down-sampling method as the ViT and uses a fast Fourier transform on the embedded patch tokens. They used a novel fusion transformer block to combine features extracted by the ViT and GFNet, taking advantage of information from both the spatial and frequency domains. Finally, voting among the predicted labels of slices produced each MRI's final prediction. Inter-slice relations are necessary, as mentioned. This method also loses temporal information between slices.

As the ViT models are developed for 2D images, applying the ViT-based models to 3D MRI scans is difficult. [27] proposed two different models with the ability to process 3D MRI scans. The first one is a 3D version of ViT (Voxel Vision Transformer: VViT) with a new 3D patch embedding method. In this model, the whole MRI voxel feeds the model input. VViT used non-overlapping 3D patches with a fixed size of 50×50×50, while vanilla ViT uses 2D patches with a size of 32×32. The second model takes advantage of the CNN in the ViT by using convolutional operations with overlapping for patch embedding (tokenization). The Convolutional Voxel Vision Transformer (CVVT) was a new convolutional patch embedding module with a series of convolution blocks and a smaller fully connected layer. They claim that when trained on insufficient data, the model does not generalize well. Despite having lower classification accuracy than most existing CNN-based models, CVVT has a lot of potential because of the idea of communicating features from different regions.

III. PROPOSED METHOD

The proposed method consists of three main steps: preprocessing; independent feature extraction from the slices using Vit; and binary classification using a bi-LSTM model to encode feature sequences. Fig. 3 summarized the proposed approaches.

*A. Preprocessing*

The proposed method for preprocessing MRI images involves skull stripping, which removes non-brain tissues to isolate and extract the brain region. This step improves accuracy and reliability by eliminating irrelevant information and reducing noise. Normalization and registration processes ensure consistent intensity values and data alignment, crucial for MRI analysis and comparison. The Montreal Neurological Institute (MNI) space is used for this purpose in the proposed method. We accomplished our goals with the help of MATLAB's Computational Anatomy Toolbox (CAT12) and the Statistical Parametric Mapping 12 tool (SPM12) [28].

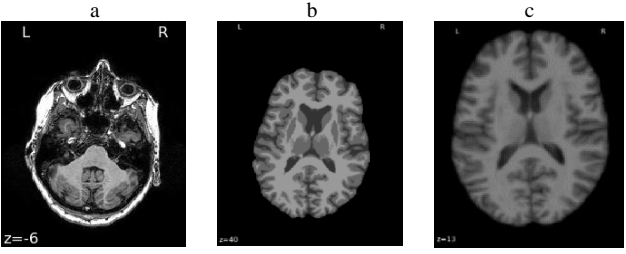

Fig. 1. Axial MRI slices, a) Original MRI, b) removed skull, and c) segmented scan warped to MNI space

*B. Feqture extraction*

Let's start with a detailed breakdown of the architecture of a vision transformer.

- **Image patching:** The input image is divided into a grid of non-overlapping, fixed-size patches. Each patch consists of a fixed number of pixels and functions as a token. (Fig. 2: A). The size of the patches can vary, but $16 \times 16$ pixels is a common option. If the input image has dimensions $H \times W \times C$ (height, width, and number of channels), the resulting grid will have dimensions $(H/P) \times (W/P)$, representing the number of patches.

- **linear embeddings:** Patches are converted into a sequence of 1D embedding vectors (Fig. 2: B). The image becomes a token sequence after flattening. $\mathbf{x} \in \mathbb{R}^{H \times W \times C}$ the image is flattened into a sequence of 2D patches $\mathbf{x}_p \in \mathbb{R}^{N \times (P^2 \cdot C)}$. where (H; W) is the image size, C is the number of color channels, (P; P) is the size of each image patch, and $N = HW/P^2$ is the number of patches.

- **Patch embeddings:** Flattened patches are linearly projected to lower-dimensional vectors, called token embeddings. These embeddings record patch content. Each patch is vectorized and projected linearly into tokens per:

$$\hat{\mathbf{x}} = [x_{class}, x_1 \mathbf{E}, x_2 \mathbf{E}, \dots, x_N \mathbf{E}], \mathbf{E} \in \mathbb{R}^{CP^2 \times D} \quad (1)$$

- **Positional embeddings:** To provide spatial information, the token embeddings are augmented with learnable positional embeddings. These embeddings encode each token's relative position in the sequence. A positional embedding known as $\mathbf{E}_{pos}$ is added to the token as follow:

$$\mathbf{z}_0 = \hat{\mathbf{x}} + \mathbf{E}_{pos}, \mathbf{E}_{pos} \in \mathbb{R}^{(N+1) \times D} \quad (2)$$

- **Transformer blocks:** The vectors are then passed through a series of transformer blocks. Each transformer block has two sub-layers: a multi-head self-attention (MSA) layer and a feed-forward layer (MLP) with Layer-Norm (LN) (Fig. 2: C). This operation is calculated per:

$$\mathbf{z}'_\ell = \text{MSA}(\text{LN}(\mathbf{z}_{\ell-1})) + \mathbf{z}_{\ell-1}, \quad \ell = 1 \dots L \quad (3)$$

$$\mathbf{z}_\ell = \text{MLP}(\text{LN}(\mathbf{z}'_\ell)) + \mathbf{z}'_\ell, \quad \ell = 1 \dots L \quad (4)$$

$$\mathbf{y} = \text{LN}(\mathbf{z}^0_L) \quad (5)$$

- **Classification head:** A classification head is then applied to the output of the transformer blocks. This is a linear layer that outputs the input image's class probabilities (Fig. 2: D).

While ViT models perform well on two-dimensional data, the limited amount of data available from 3D MRIs makes it challenging to expect effective results. We normalized a 3D

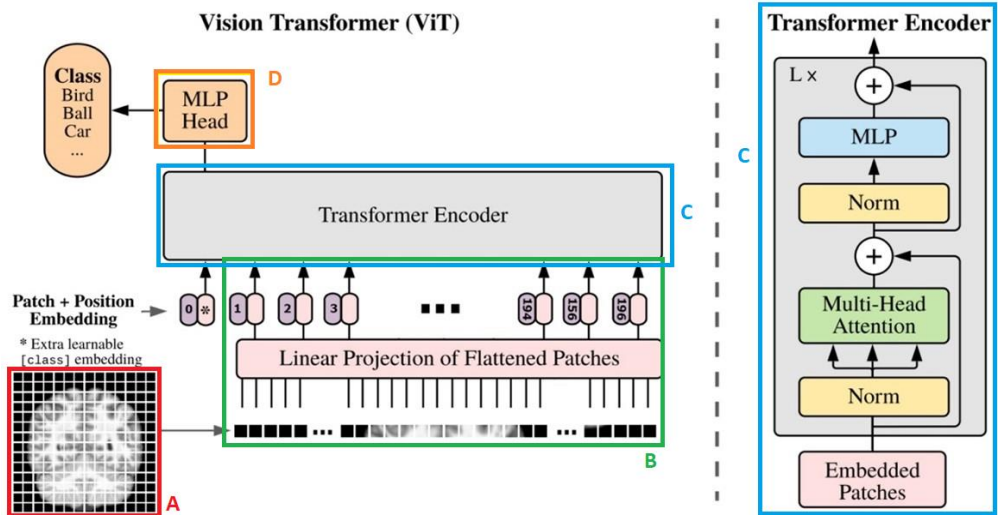

Fig. 2. Overview of the ViT model

MRI in the axial plane and cut it into 2D slices so that a pre-trained 2D network model could be used for transfer learning. With this method, we can learn more about neural networks and improve transfer learning. Brain MRI data were extracted independently from each slice in the axial plane using the Vision Transformer (ViT) method. In neuroimaging analysis, the axial plane refers to the cross-sectional view of the brain. By applying the ViT model to each individual slice, the method identifies pertinent visual patterns and characteristics within each slice. This method permits the extraction of slice-specific characteristics, enabling a comprehensive local representation of the data. Utilizing the power of deep learning and the attention mechanisms employed by ViT, the method is able to effectively identify and highlight important information within each slice, thereby facilitating subsequent analysis and classification tasks. We will ignore the classification head layer and take the Pooler output, a linear projection of the last layer's feature vectors, since we use the pre-trained ViT as a feature extractor.

*C. Bi-LSTM for sequence data classification*

The acquired MRI slice features were used as the input to the Bi-LSTM network for the classification of Alzheimer's disease (AD). Bi-LSTM is derived from the network for long-term, short-term memory (LSTM) [29]. LSTM is a special type of recurrent neural network (RNN). When learning long sequences, RNNs typically face gradient disappearance and gradient explosion issues [30]. LSTM was created to solve this issue by enhancing its internal structure [31]. It is capable of capturing the data's time dependence and long-term dependence [32]. LSTM's gating mechanism can effectively prevent gradient vanishing and improve the model's training efficiency and accuracy[33].

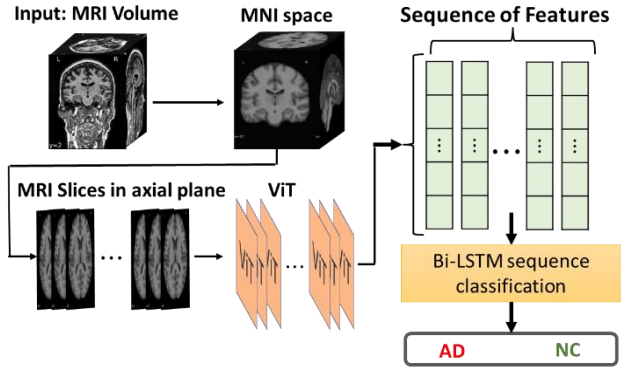

Fig. 3. Summarization of proposed model.

## IV. RESULTS

This section presents experimental results and compares the proposed method to relevant literature. The variants of the ViT and Bi-LSTM models are described in TABLE I. and TABLE II.

We have reported the classification performance of the proposed method on ADNI data collections (ADNI1: Complete 1Yr 1.5T) [34], [35]. Accuracy, precision, F-score, and recall(sensitivity) are the four metrics utilized to evaluate each classification model's effectiveness. For training and testing, we used the K-fold cross-validation protocol with K set to 10. Overfitting to training data is reduced by K-fold cross-validation. TABLE I. shows a comparison between our best results and some relevant studies that are already in literature.

TABLE I. DETAILS OF MODEL VARIANTS

| Parameter name | Values |
|---|---|
| ViT | |
| Model | vit-base-patch16-224-in21k |
| Trained with | ImageNet-21k (14 million images, 21,843 classes) at resolution 224 × 224 |
| Path resolution | 16 × 16 |
| Layers | 12 |
| Hidden size D | 768 |
| MLP size | 3072 |
| Heads | 12 |
| Params | 86M |

TABLE II. DETAILS OF MODEL VARIANTS

| Parameter name | | Values |
|---|---|---|
| Layers | | 6 |
| Hidden Units | | 64 |
| Epoch | | 100 |
| Batch Size | | 25 |
| Adam Optimizer | learning rate | 1e-4 |
| | $\beta_1$ | 0.9 |
| | $\beta_2$ | 0.999 |
| | $\epsilon$ | 1e-07 |
| Loss Function | | Sparse Categorical Cross entropy |
| Dropout | | 0.15 |
| Activation | | tanh |
| Input Dimension | | 50 × 2048 |
| Output Dimension | | #classes (2/3) |
| Params | | 538,515 |

TABLE III. STATE-OF-THE-ART BINARY CLASSIFICATION PERFORMANCE COMPARISON

| work | Image scans (NC/AD) | Method | NC/AD classification | | |
|---|---|---|---|---|---|
| | | | ACC (%) | SEN (%) | SPE (%) |
| [36] | 429/-/858 | 3D CNN | 90.3 | 82.4 | 96.5 |
| [37] | 119/-/97 | 3D DenseNet | 88.9 | 86.6 | 90.8 |
| [38] | 330/-/299 | 3D CNN | 93.2 | 95.0 | 89.8 |
| [39] | 324/-/319 | self-attention | 98.0 | 97.7 | 98.2 |
| [40] | 457/-/346 | 3D ResNet | 94.00 | - | - |
| [41] | 209/-/188 | 2.5D CNN | 79.90 | 84.00 | 74.80 |
| Our Method | 705/-/476 | ViT-Bi-LSTM | 95.678 | 95.5 | - |

Diverse data samples presented a problem when comparing our study to those of others. For instance, the selection of all the data in TABLE III. may be made in accordance with personal preference. The second column of these tables shows that not only are the samples unequal, but the data may not overlap. To this end, we used already-collected data from ADNI. The name of the data collection is ADNI1: Complete 1Yr 1.5T. So, the later researcher can do a fair comparison by choosing the same data collection.

Work [36] achieved an accuracy of 90.3%, with a sensitivity of 82.4% and a specificity of 96.5%. This suggests that the method performed reasonably well in correctly classifying both positive (AD) and negative (NC) cases, with a higher accuracy in identifying NC. Work [37] achieved an accuracy of 88.9%, a sensitivity of 86.6%, and a specificity of 90.8%. These results indicate a good overall performance in correctly identifying both AD and NC cases, although with slightly lower accuracy compared to Work [36]. Work [38] achieved an accuracy of 93.2%, with a sensitivity of 95.0% and a specificity of 89.8%. These findings suggest a strong performance in correctly identifying both AD and NC cases, with a higher accuracy in identifying AD. Work [39] achieved an accuracy of 98.0% and a sensitivity of 97.7%. However, the specificity value is not provided in the table, so we cannot determine the exact performance in identifying NC cases. It is advisable to refer to the original source or additional information to obtain the accurate specificity value for this study. Work [40] achieved an accuracy of 94.0%, but the sensitivity and specificity values are not available in the table. Therefore, it is challenging to assess the complete performance of this method based solely on the accuracy metric. Work [41] achieved an accuracy of 79.9%, a sensitivity of 84.0%, and a specificity of 74.8%. These results suggest a relatively lower performance compared to other methods, with a higher rate of false positives (incorrectly identifying NC as AD) and false negatives (incorrectly identifying AD as NC). Regarding "Our Method," it achieved an accuracy of 95.678% and a sensitivity of 95.5%.

Overall, based on the available information, our proposed method shows promise and appears to be a competitive approach for NC/AD classification. All the available information from the table suggests that some methods demonstrated good performance in accurately classifying NC and AD cases, while others showed lower performance or had missing specificity values.

V. CONCLUSION

Using deep learning techniques to analyze and classify AD patients, the use of ViT and bi-LSTM illustrates a novel approach to the diagnosis of AD. ADNI (Alzheimer's Disease Neuroimaging Initiative) data were used to evaluate the performance of the proposed model. The results indicate that the proposed method outperforms other deep learning models evaluated in the literature, demonstrating its accuracy in classifying AD patients. Early and accurate diagnosis of Alzheimer's disease is essential for timely treatment and improved disease management, potentially slowing the disease's progression and enhancing the quality of life for those affected. Overall, the results indicate that the proposed method combining ViT and bi-LSTM holds promise as a reliable and effective diagnostic tool for AD. Further validation on larger and more diverse datasets, as well as comparison with other well-established methods, would strengthen its potential impact and pave the way for future research in this field.

ACKNOWLEDGMENT

Data used in the preparation of this article were obtained from the Alzheimer's disease Database Initiative (ADNI) database. A complete listing of ADNI investigators can be found at: http://adni.loni.usc.edu/wp-content/uploads/how_to_apply/ADNI_Acknowledgement_List.pdf.